\documentclass[twocolumn]{aastex62}

\usepackage{graphicx,amsmath,amssymb,amstext,natbib}
\usepackage{here}
\usepackage{epsf}
\usepackage{epstopdf}
\usepackage{xspace}

\usepackage[normalem]{ulem} 
\usepackage{cancel} 

\usepackage{color}

\newcommand{\um}{\ensuremath{\mu\rm{m}}\xspace}
\newcommand{\kms}{\ensuremath{\rm{km\,s}^{-1}}\xspace}
\newcommand{\alphaco}{\ensuremath{\alpha_{\rm{CO}}}\xspace}

\newcommand{\uJy}{\ensuremath{\mu\rm{Jy}}\xspace}
\newcommand{\lfir}{\ensuremath{L_{\rm{FIR}}}\xspace}
\newcommand{\lir}{\ensuremath{L_{\rm{IR}}}\xspace}

\newcommand{\Mstar}{\ensuremath{M_{\rm{star}}}\xspace}

\newcommand{\Mht}{\ensuremath{M_{\rm{H}_2}}\xspace}

\newcommand{\Msol}{\ensuremath{\rm{M}_\odot}\xspace}

\newcommand{\ML}{\ensuremath{M/L}\xspace}

\newcommand{\fht}{\ensuremath{f_{\rm{H}_2}}\xspace}
\newcommand{\tdep}{\ensuremath{t_{\rm{dep}}}\xspace}

\newcommand{\reff}{\ensuremath{r_\mathrm{eff}}\xspace}

\newcommand{\etal}{et~al.\xspace}
\newcommand{\arc}{\ensuremath{''}\xspace}

\shortauthors{J. S. Spilker, et al.}
\shorttitle{Inside-Out Quenching at $z$ $\sim$ 2}

\begin{document}

\defcitealias{spilker16b}{S16}

\title{\uppercase{Evidence for Inside-Out Galaxy Growth and Quenching of a\\\MakeLowercase{z} $\sim$ 2 Compact Galaxy From High-Resolution Molecular Gas Imaging}}

\correspondingauthor{Justin S. Spilker}
\email{spilkerj@gmail.com}

\author[0000-0003-3256-5615]{Justin S. Spilker}
\affiliation{Department of Astronomy, University of Texas at Austin, 2515 Speedway, Stop C1400, Austin, TX 78712, USA}

\author[0000-0001-5063-8254]{Rachel Bezanson}
\affiliation{Department of Physics and Astronomy and PITT PACC, University of Pittsburgh, Pittsburgh, PA 15260, USA}

\author[0000-0001-6065-7483]{Benjamin J. Weiner}
\affiliation{Steward Observatory, University of Arizona, 933 North Cherry Ave., Tucson, AZ 85721, USA}

\author[0000-0001-7160-3632]{Katherine E. Whitaker}
\affiliation{Department of Physics, University of Connecticut, 2152 Hillside Road, Unit 3046, Storrs, CT 06269, USA}
\affiliation{Cosmic Dawn Center (DAWN), Niels Bohr Institute, University of Copenhagen / DTU-Space, Technical University of Denmark}

\author[0000-0003-2919-7495]{Christina C. Williams}
\altaffiliation{NSF Fellow}
\affiliation{Steward Observatory, University of Arizona, 933 North Cherry Ave., Tucson, AZ 85721, USA}

\begin{abstract}

We present high spatial resolution imaging of the CO(1--0) line from the Karl G. Jansky Very Large Array (VLA) of COSMOS\,27289, a massive, compact star forming galaxy at $z=2.234$. This galaxy was selected to be structurally similar to $z\sim2$ passive galaxies. Our previous observations showed that it is very gas-poor with respect to typical star-forming galaxies at these redshifts, consistent with a rapid transition to quiescence as the molecular gas is depleted. The new data show that both the molecular gas fraction $\fht \equiv \Mht/\Mstar$ and the molecular gas depletion time $\tdep \equiv \Mht/$SFR are lower in the central 1--2\,kpc of the galaxy and rise at larger radii $\sim$2--4\,kpc. These observations are consistent with a scenario in which COSMOS\,27289 will imminently cease star formation in the inner regions before the outskirts, i.e. inside-out quenching, the first time this phenomenon has been seen via observations of molecular gas in the high-redshift universe. We find good qualitative and quantitative agreement with a hydrodynamical simulation of galaxy quenching, in which the central suppression of molecular gas arises due to rapid gas consumption and outflows that evacuate the central regions of gas. Our results provide independent evidence for inside-out quenching of star formation as a plausible formation mechanism for $z\sim2$ quiescent galaxies.

\end{abstract}


\section{Introduction} \label{intro}

Massive quiescent galaxies have been identified early in the history of the universe, increasing rapidly in number density since $z\sim4$ \citep[e.g.][]{kriek06,whitaker10,cassata13,straatman14}. These early quiescent galaxies are in some ways strikingly different from local giant elliptical galaxies, with typical sizes $\sim$5$\times$ smaller at a given mass \citep[e.g.][and references therein]{vanderwel14}. While similarly massive,  compact galaxies are very rare in the nearby universe, the central stellar densities in the inner 1\,kpc are very similar at $z\sim2.3$ and $z\sim0$ \citep{bezanson09,belli14a}, implying an inside-out formation scenario in which subsequent growth since $z\sim2$ has mostly occurred at larger radii \citep{williams17,lee18}. This late-time evolution is generally consistent with size growth due to gas-poor minor mergers, with quiescent galaxies remaining approximately virialized over cosmic time \citep[e.g.][]{trujillo11,newman12,bezanson13}.

The formation mechanism(s) of these early quiescent galaxies is still unclear. Most theoretical simulations that successfully form passive galaxies that are comparably compact to the observed population require gas-rich dissipative processes, including mergers, counter-rotating cosmological accretion, or disk instabilities \citep[e.g.,][]{naab09,feldmann15,wellons15,ceverino15,zolotov15,pandya17}. High gas fractions permit central growth as the dissipative process (whatever its origin) redistributes angular momentum and drives gas inward. These simulations predict a compact star-forming phase immediately prior to the quenching of star formation, caused by some combination of gas consumption and heating/winds due to galactic feedback. Given the structural similarity, these galaxies need only cease star formation in order to match observed properties of $z\sim2$ quiescent galaxies.

Significant numbers of these compact star-forming galaxies (SFGs) have now been found and characterized \citep[e.g.,][]{stefanon13,barro13,barro14,williams14,vandokkum15}. Such compact SFGs host active galactic nuclei (AGN)  more frequently than similar-mass extended star-forming galaxies at these redshifts \citep{barro13,kocevski17}, suggesting that AGN may play a role in the evolution of this population.  Dynamical models suggest that stars make up the vast majority of the mass within the central several kiloparsecs, with little room for other components \citep{barro14,vandokkum15,wisnioski18}. This has recently been confirmed by VLA and Atacama Large Millimeter Array (ALMA) observations, which find that compact SFGs are remarkably gas-poor compared to similar-mass and -SFR galaxies at these redshifts \citep{spilker16b,barro17,popping17,tadaki17,talia18}. Combined with their high SFRs, these observations provide independent evidence that compact SFGs are indeed likely to become quiescent on short timescales.

On spatially resolved scales, most studies find that massive high-redshift galaxies build their stellar mass in an inside-out fashion. This conclusion is based on resolved radial gradients in the specific SFR (sSFR $\equiv$ SFR/\Mstar), with rising sSFR profiles indicative of faster growth in the outskirts of galaxies compared to the centers \citep[e.g.,][]{wuyts13,tacchella15,tacchella18,nelson16}. Whether these results also hold for SFGs specifically selected to be compact is unclear. The SFR profiles have been derived from H-$\alpha$ and rest-UV imaging, which are susceptible to dust obscuration and AGN contribution to the H-$\alpha$ flux, both of which are particularly acute features of compact SFGs. \citet{barro16} presented very high angular resolution ALMA imaging of 6 compact SFGs at $z\sim2.5$ and found that the SFR profiles were more concentrated than the stellar mass profiles, implying that those galaxies are forming compact central stellar bulges, consistent with the gas-rich dissipative processes seen in simulations.

An alternative approach to studying the growth of galaxies comes from high-resolution observations of molecular gas, the fuel for star formation, either from tracer molecules such as carbon monoxide (CO) or from long-wavelength dust emission. Compared to measurements at shorter wavelengths, gas tracers have the advantage of being little-influenced by the presence of dust or AGN. While not free of systematic uncertainties, the systematics are at least different from and arguably better understood than those that affect rest-optical SFR profile measurements, providing an independent view of the assembly of these galaxies.

Here we present high-resolution VLA observations of CO(1--0) in one such compact SFG, COSMOS\,27289. Typical of this population of galaxies, COSMOS\,27289 has a stellar mass $\Mstar \sim 1.3 \times 10^{11}$\,\Msol, SFR\,$\sim400$\,\Msol/yr, and a compact size $\reff \sim 2.3$\,kpc. The galaxy is undetected in X-ray imaging. Our previous lower-resolution VLA observations of this object found a very low gas fraction $\fht \approx 0.1$ and extremely short depletion time $\tdep \approx 30$\,Myr (\citealt{spilker16b}, hereafter \citetalias{spilker16b}). Many compact SFGs exhibit unexpectedly low H-$\alpha$ line widths, given their large masses and compact sizes, or equivalently, stellar masses larger than simple dynamical mass estimates \citep{vandokkum15}. COSMOS\,27289 is an extreme example of this, with full-width at half-maximum line widths of $\sim130$\,\kms observed in H-$\alpha$ and an even narrower CO(1--0) line width, $\sim$60\,\kms (\citealt{vandokkum15}, \citetalias{spilker16b}). This very narrow emission implies that COSMOS\,27289 must be very nearly face-on.\footnote{There are other possible explanations. The gas could be more compact with respect to the stars, such that the line width need not reflect the total stellar mass of the galaxy. Our new observations show that this is not the case. Alternatively, the stellar mass could be drastically overestimated by two orders of magnitude, although this seems unlikely given the extensive multiwavelength photometry available in the COSMOS field.} This offers us the chance to observe the radial distributions of gas, dust, and stars in this galaxy with little uncertainty due to the effects of inclination.

In the remainder of this work we describe our new VLA observations and present evidence of a central suppression of the gas fraction in COSMOS\,27289 consistent with expectations for the imminent inside-out quenching of star formation. Section~\ref{data} describes the new and archival observations we use in this work, our image- and visibility-based radial profile analysis techniques, and our assumptions for converting observables to physical quantities. Section~\ref{results:fgashole} presents our main observational findings, Section~\ref{assume} examines the limitations of our assumptions in the context of our results, and we compare our findings to simulations in Section~\ref{results:sim}. We place COSMOS\,27289 in context with other galaxy populations in Section~\ref{results:context}, and conclude in Section~\ref{conclusions}. Throughout, we assume a flat $\Lambda$CDM cosmology ($H_0 = 67.7$\,\kms\,Mpc$^{-1}$, $\Omega_m = 0.307$; \citealt{planck16}).

\section{Data and Analysis} \label{data}

\begin{figure*}[htb]
\includegraphics[width=\textwidth]{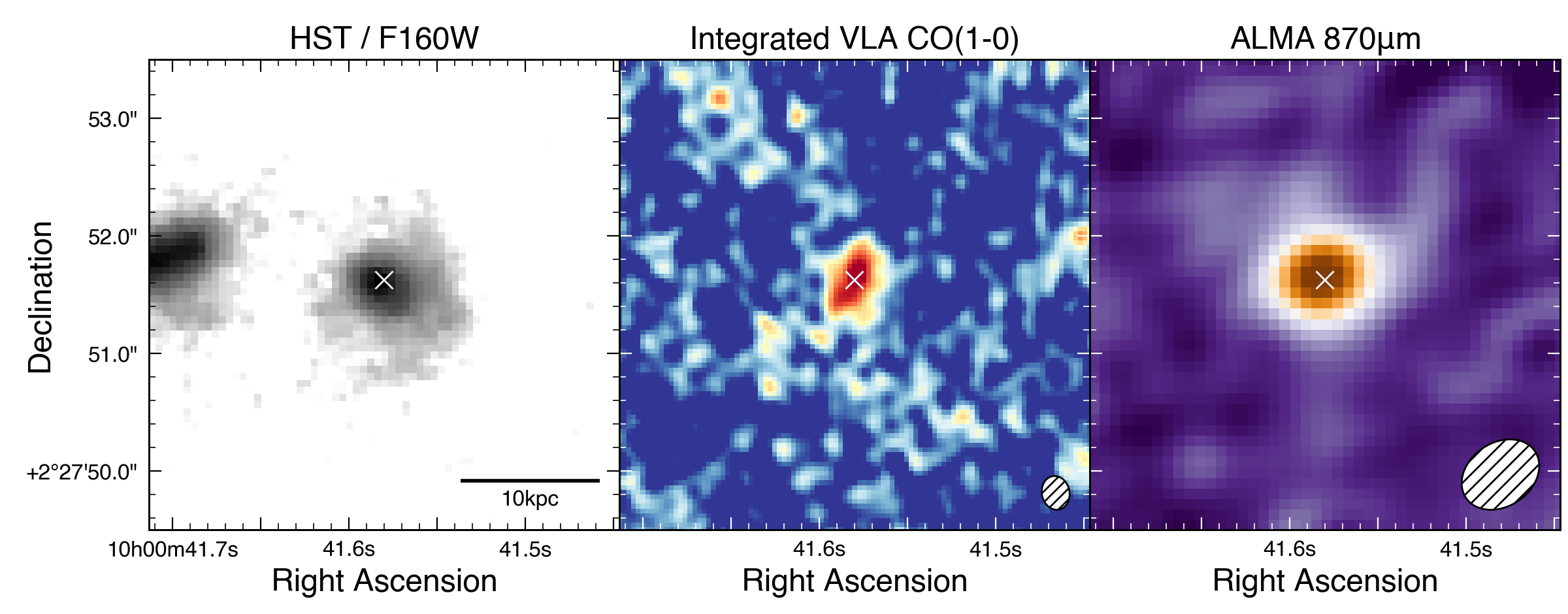}
\caption{
Overview of the data products used in this work.
\textit{Left: } HST WFC3/F160W image, tracing rest-frame $\sim$500\,nm. A centrally-peaked profile with additional low surface brightness emission is apparent. The peak of this image is marked with a white $\times$, and is repeated in the other two panels. All images peak at the same location to within their mutual uncertainties.
\textit{Center: } Integrated VLA CO(1--0) image, made from combining previously-obtained C array configuration data with new B array data. This image integrates over 125\,\kms, and captures $\approx$99\% of the emission seen in the previous data.
\textit{Right: } Archival ALMA 343\,GHz image, tracing rest-frame 270\,\um. For both the VLA and ALMA data, the ellipses at lower right indicate the sizes of the synthesized beams.
}\label{fig:images}
\end{figure*}

We make use of new and archival data from the VLA, targeting CO(1--0) emission, as well as archival HST and ALMA data. The imaging data we use in our subsequent analysis are shown in Figure~\ref{fig:images}.

\subsection{VLA Observations}
In previous work, we observed the CO(1--0) emission line in COSMOS 27289 in project 16A-203 \citepalias{spilker16b}. These observations, using the VLA's C array configuration, yielded $\approx0.8$'' spatial resolution, and indicated that the CO emission was not significantly extended on scales of $\approx$7\,kpc.  We combine those data with newly-obtained observations using the B array configuration, reaching an effective spatial resolution of $\approx$0.25'' ($\approx$2\,kpc). Further details of the lower-resolution data are available in \citetalias{spilker16b}.

The B array observations were taken in October 2017 in project 17B-223 (PI: Spilker). We exactly replicated the correlator configuration of the previous data for ease of data combination, resulting in 2\,GHz of continuous bandwidth using the 8-bit samplers, subdivided into sixteen 128\,MHz dual-polarization basebands with 1\,MHz channelization. As in the previous data, the quasar 3C286 served as bandpass and absolute flux calibrator, while quasar J0948+0022 served as the complex gain calibrator. The observations were executed in three 5-hour tracks, with a total on-source time of 7.0 hours in the B array. To test whether the new higher-resolution data have resolved out diffuse CO emission, we measured the integrated CO line flux from the C- and B-array data separately. We found no significant difference in the line fluxes, which indicates that the B-array data capture the entirety of the CO emission alone. 

We combined the data from the two array configurations by rescaling the data weights for each baseline from each observing session based on the scatter in the visibilities on that baseline. Because COSMOS 27289 is very faint, this is equivalent to more sophisticated time-dependent differencing methods, as each measured visibility contains virtually no signal. We then imaged the data using natural weighting, integrating over 125\,\kms centered on the CO(1--0) systemic velocity determined from the previous low-resolution data. This binning in velocity contains $\approx$99\% of the emission in the previous data; the CO(1--0) line in this source is quite narrow. The combined C and B array data reach a sensitivity of 34\,\uJy/beam in the 125\,\kms channel width we use in our analysis. The resulting image is shown in Figure~\ref{fig:images}, and reaches a peak signal-to-noise (S/N) of $\sim$8 in integrated CO emission. We experimented with finer channelization of the data to probe for, e.g., resolved kinematics, but the low S/N of the data preclude robust conclusions on this point.

\subsection{Archival ALMA and HST Data}

We make use of archival data from ALMA and HST, tracing dust heated by recent star formation and stellar light, respectively.

COSMOS\,27289 was observed in the HST WFC3/F160W filter as part of the CANDELS program, with multi-band photometry collated as part of the 3D-HST program \citep{koekemoer11,brammer12,skelton14,momcheva16}. At $z=2.234$, this filter probes rest-frame $\sim$500\,nm, expected to arise mostly from the bulk of the existing stellar mass within the galaxy, and the data reach a spatial resolution of $\approx$0.18\arc. While this galaxy was also observed in the WFC3/F125W filter in the CANDELS program, this filter probes only rest-frame $\sim$390\,nm. The F125W filter bandpass straddles the 4000\,\AA break and the light in this image is thus some weighted average of the older and younger stellar populations, while the F160W image is a better tracer of the stellar mass spatial distribution. This source is also weakly detected in ACS/F814W imaging \citep{scoville07}, which samples the rest-ultraviolet. However, the S/N in this band is low, and in any case the unobscured star formation is only a minor contribution to the total SFR (UV-based SFR $\approx$ 8\,\Msol/yr, IR-based SFR $\approx$ 400\,\Msol/yr; see Section~\ref{conv}).

We also make use of archival ALMA 343\,GHz imaging of COSMOS\,27289 from project 2015.1.00137.S. We downloaded, re-reduced, and re-imaged these data using the standard ALMA pipeline. The ALMA data reach a spatial resolution of $\sim$0.5$\times$0.7\arc and a sensitivity of 0.14\,mJy/beam when combining the full 7.5\,GHz of observed bandwidth using natural weighting of the visibilities, and detect the source at a peak S/N of $\approx$20. While these data have lower spatial resolution than the HST or VLA data, we demonstrate below that the ALMA detection is marginally spatially resolved, allowing for (admittedly weak) constraints on the distribution of the dust emission. These data sample rest-frame $\sim$270\,\um, a somewhat longer wavelength than the typical peak of the dust SED, but also not fully on the Rayleigh-Jeans tail of the dust emission. The ALMA imaging thus somewhat traces the temperature-weighted dust mass distribution within this galaxy \citep[e.g.][]{scoville16} as well as the ongoing star formation obscured by dust.  From both the 3D-HST catalogs and analysis of archival \textit{Herschel} imaging \citepalias{spilker16b}, we estimate that $\gtrsim$95-98\% of the ongoing SFR is obscured by dust \citep[see also][]{whitaker17b}, so we expect that the ALMA data probe essentially all of the star formation in this object.

\subsection{Radial Profile Fitting}

We make use of radial surface brightness profiles in order to help interpret the spatial distribution of molecular gas, stars, and obscured star formation in COSMOS\,27289. In each case we generate an elliptical S\'{e}rsic profile model of the galaxy, deconvolved from the point spread function (PSF; HST) or the synthesized beams (VLA and ALMA). We then employ a first-order correction for the fact that the galaxy need not be perfectly represented by a single S\'{e}rsic profile, following \citet{szomoru10,szomoru12}. A more detailed description of this procedure is available in those works. Briefly, we measure the radial surface brightness profile from the model-subtracted residual images, and add this residual profile to the best-fit intrinsic (deconvolved) model profile. While the final radial profile is the sum of the \textit{deconvolved} S\'{e}rsic profile model and the PSF- or beam-\textit{convolved} residual profile, \citet{szomoru10} demonstrate that this method results in profiles that are less sensitive to the best-fit S\'{e}rsic index $n$ because the imperfections in the assumed model are re-incorporated into the final measured radial profile.

\begin{figure}[htb]
\includegraphics[width=\columnwidth]{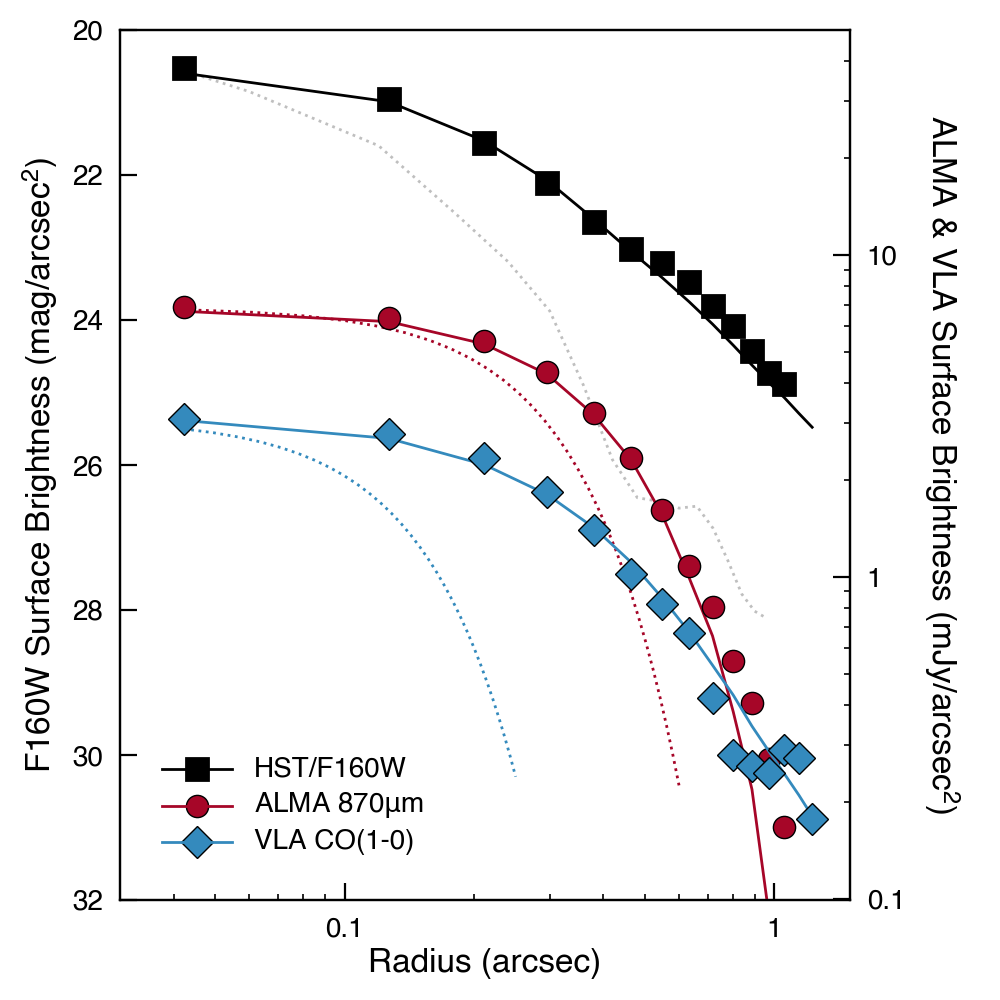}
\caption{
Measured radial profiles from the WFC3/F160W, VLA CO(1--0), and ALMA 343GHz images, along with the best-fit single S\'{e}rsic model for each. Both data and model are convolved with the PSF (HST) or synthesized beams (VLA and ALMA). Dotted lines indicate the PSF (for HST) or Gaussian synthesized beam shapes, as an indication of the measured source sizes with respect to the resolution of the data. Note that the left (HST) and right y-axes (VLA and ALMA) are distinct, with arbitrary relative scaling.
}\label{fig:dataprofs}
\end{figure}

For the HST data, we fit the WFC3/F160W image using the \texttt{GALFIT} software \citep{peng10}, following previous works by fitting a single S\'{e}rsic profile to the image \citep{vanderwel14,vandokkum15}. We use the image and PSF generated by the 3D-HST team \citep{skelton14}. We find structural parameters in excellent agreement with the values published previously for this galaxy \citep{vanderwel14,vandokkum15}. The (PSF-convolved) radial profile and convolved model are shown in Figure~\ref{fig:dataprofs}. The rest-frame 500\,nm light is dominated by a centrally-concentrated component with S\'{e}rsic index $n = 3.1 \pm 0.1$ and circularized effective radius $2.6 \pm 0.1$ kpc. 

Neither the VLA nor ALMA data is as well-resolved as the HST image of this galaxy in terms of number of resolution elements across the source. While image-based fitting routines such as \texttt{GALFIT} can in principle include the effects of the relatively large Gaussian synthesized beams, these fitting techniques may perform poorly in cases such as this, where the source is resolved into only a few independent resolution elements. Instead we turn to a visibility-based technique that fits directly to the Fourier components measured by the interferometers. This method fully incorporates the information present in the interferometric visibilities and avoids the correlated uncertainties present in the inverted images of the galaxy.

Several routines are available that can fit simple models to visibilities (e.g., \texttt{uvmodelfit} in the CASA interferometric data reduction software; \citealt{mcmullin07}), but these models are limited to those with simple analytic Fourier transforms (e.g. a Gaussian, whose Fourier transform is also a Gaussian in the $uv$ plane). In order to replicate our fitting procedure to the HST data as closely as possible, we used the \texttt{visilens} code originally developed for modeling of gravitational lensing systems in the $uv$-plane \citep{spilker16}, slightly modified for use on this unlensed source. This code uses a Bayesian Markov Chain Monte Carlo (MCMC) sampling algorithm to determine the best-fit shape parameters and their covariances. 

Because a S\'{e}rsic profile with generic index $n$ does not have a closed-form Fourier transform, we instead generate a model image based on a set of trial S\'{e}rsic parameters sampled at a resolution several times higher than the synthesized beam size. This model image is then Fourier transformed, sampled at the $uv$ coordinates of the VLA or ALMA data, and compared to the data with a standard $\chi^2$ figure of merit based on the uncertainty of each visibility. The result is effectively a best-fit model deconvolved from the dirty beam of the data (itself determined by the $uv$ coverage during the observation). We measure the residual profile from the Fourier transform of the residual (data minus model) visibilities; as with the HST fitting procedure, this residual profile is still convolved with the dirty beam.

We note that the peak (centroid) emission of both the CO and dust continuum emission is cospatial with the F160W centroid to less than a synthesized beam. While the ellipticity and position angle of the best fit S\'{e}rsic profiles are not well constrained for the VLA or especially ALMA data, they are at least statistically consistent with the best-fit values from HST. We thus measure all radial profiles in identical elliptical apertures centered on the peak of the F160W emission, with ellipticity and position angle of the apertures determined from the model fit to the F160W image.

The convolved radial profiles of the best-fit models and data are shown in Figure~\ref{fig:dataprofs}, along with the profiles from the synthesized beams, which are Gaussian by construction \citep{hogbom74}. For both the VLA and ALMA data, little significant structure is seen in the residuals. Reassuringly, our visibility-based fitting procedure also determines that the source is moderately resolved in the ALMA data. We determine the uncertainties on each profile by drawing random samples from the MCMC chains (for VLA and ALMA) or by generating artificial profiles using the \texttt{GALFIT}-derived uncertainties assuming they are normally distributed and not covariant.

\subsection{Converting Observed to Physical Quantities}\label{conv}

We make very simple assumptions to convert each surface brightness profile into physical units. Namely, we assume a single, global conversion factor for each component -- a single mass-to-light ratio for the F160W image, a single \alphaco conversion factor for the CO(1--0) data, and a single 270\,\um-to-SFR conversion for the ALMA data. While we expect that these assumptions do not reflect reality, we argue below that deviations in each case would exaggerate the main effect we observe -- namely, a relative dearth of molecular gas in the central kiloparsecs that leads to suppressed gas fractions and depletion times compared to the outskirts of the galaxy.

For the F160W profile, we assume a single mass-to-light ratio using the total measured flux in this band and the stellar mass from full UV-to-IR SED fitting from the 3D-HST catalog, $1.3 \times 10^{11}$\,\Msol. As previously mentioned, insufficient high-resolution data are available to attempt a more complex radially-varying mass-to-light approach. For the CO(1--0) data, we assume a single CO-H$_2$ conversion factor $\alphaco = 1$\,\Msol (K\,\kms pc$^2$)$^{-1}$ as in our previous work (\citetalias{spilker16b}; hereafter we neglect the units of \alphaco). As we argued in that work, the high SFR and compact size of COSMOS\,27289 both point towards relatively low values of \alphaco. Finally, we convert the rest-frame 270\,\um profile to a radial SFR profile by assuming a simple linear conversion between the 270\,\um flux density and the total SFR measured from SED fitting to the far-IR photometry. This source was detected by the \textit{Herschel}/PACS and SPIRE instruments in addition to the ALMA data used here, which constrain its IR luminosity and dust-obscured SFR well. We fit a standard modified blackbody to this photometry to measure \lir and translate this to an obscured SFR using standard conversions \citep[e.g.][]{kennicutt12}, yielding SFR $\approx$400\,\Msol/yr.

We assess the potential impacts of these assumptions on our principle findings in Section~\ref{assume}.

\section{Results and Discussion} \label{results}

\subsection{Depletion of Molecular Gas in the Central Kiloparsecs} \label{results:fgashole}

We begin our analysis of the distributions of molecular gas and stellar mass with a simple two-dimensional approach. In order to make a fair comparison, we convolve the higher-resolution HST data with a Gaussian kernel in order to match the $\sim$0.25'' resolution of the VLA data. We do not perform a similar exercise for the even lower-resolution ALMA data because the 870\,\um image only marginally resolves the source. Because the original HST data had $\sim$2$\times$ better resolution than the VLA data, the PSF of the convolved map is dominated by the Gaussian kernel instead of the non-Gaussian (and non-azimuthally symmetric) features of the HST PSF. We then simply scale the CO(1--0) and F160W images to maps of the molecular gas and stellar mass, respectively, as discussed in Section~\ref{conv}.

\begin{figure}[htb]
\includegraphics[width=\columnwidth]{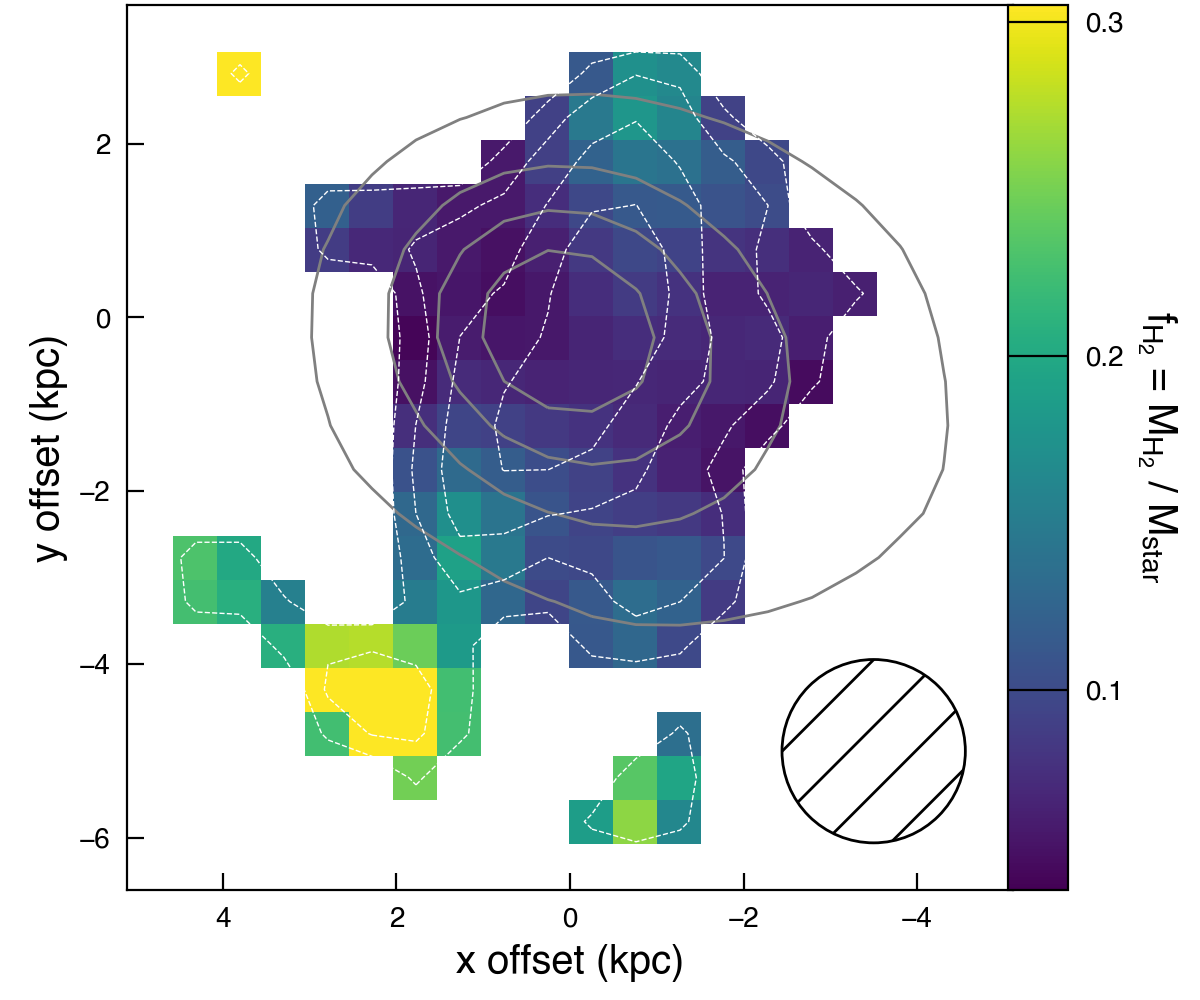}
\caption{
Two-dimensional map of the gas fraction \fht in COSMOS\,27289. The F160W image used as a proxy for the stellar mass has been convolved with a Gaussian kernel to match the resolution of the CO(1--0) data used as a proxy for the molecular gas. Contours of this smoothed image are overplotted in solid grey at 0.2, 0.4, ... times the peak value, and contours of the CO image are shown in dashed white at the same intervals. The 0.25'' ($\approx$2.0\,kpc) resolution of this map is illustrated with the hatched ellipse at lower right. The image is masked for pixels in which the CO emission is detected at $<$2$\sigma$. 
}\label{fig:fgasmap}
\end{figure}

The resulting map of the gas fraction \fht is shown in Figure~\ref{fig:fgasmap}, where we have masked pixels without significant CO emission. Again, the peaks of the stellar and CO emission are coincident to less than a resolution element. We find quite low values of the gas fraction in the central regions of the galaxy which rise out to radii of $\sim$2-4\,kpc. At larger radii the CO emission falls below our detection limit. We note that the non-azimuthally symmetric structure seen in Figure~\ref{fig:fgasmap} at radii $\gtrsim4$kpc appears to be real. Given the convolved stellar mass map and the CO sensitivity, it would have been possible to detect CO to the outermost contour in Figure~\ref{fig:fgasmap} if the gas fraction reached uniform values of $\sim$0.15-0.2 as it does in some regions. We also note that the small region of apparently high gas fraction $\sim$5\,kpc southeast of the peak of the emission corresponds to a $\sim$3$\sigma$ detection of CO; this region does not appear as a distinct clump in the HST image. If real, this may be a gas-rich clump within COSMOS\,27289 or a lower-mass companion galaxy. Deeper data would be required to verify the reality of this emission.

We also investigate the gas fraction and depletion time using a radial profile analysis. In contrast to the two-dimensional analysis, in which we convolved the high-resolution HST data to the lower resolution of the VLA data, one benefit of this radial analysis is that each component is represented by a model deconvolved from the PSF or interferometric beam, including a first-order correction by re-incorporating the residual emission not well-fit by the model. The uncertainties in each profile are propagated through to the profiles of the gas fraction and depletion time; these uncertainties by definition incorporate the effects of the differing resolutions of each dataset.

\begin{figure*}[htb]
\includegraphics[width=\textwidth]{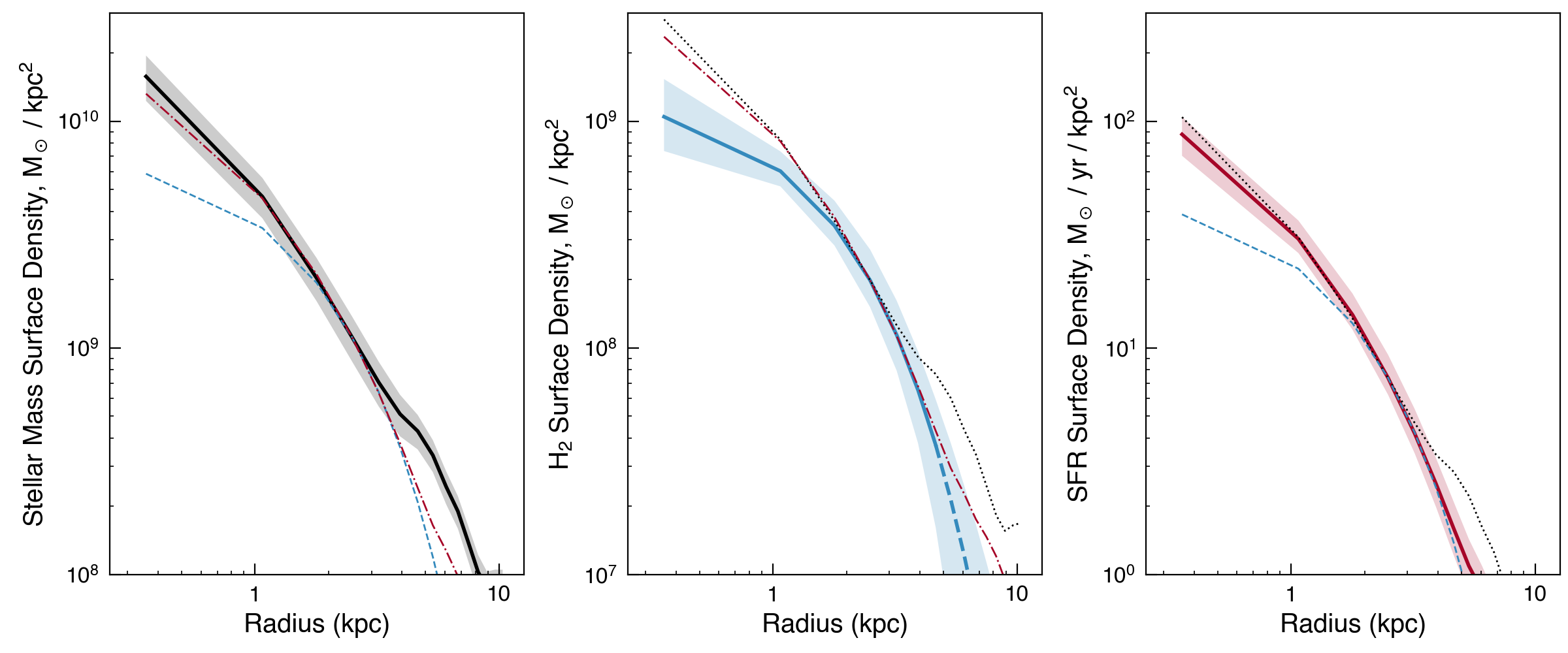}
\caption{
Radial profiles of the stellar mass, molecular gas mass, and SFR surface densities of COSMOS\,27289, plotted with the same logarithmic dynamic range ($\sim$2.5dex) to facilitate comparison. In each panel, we also plot the other two radial profiles, scaled to match at the stellar effective radius. The stellar mass is shown with a black dotted line, molecular gas with blue dashed, and SFR with a red dot-dashed line. The stellar mass is clearly more centrally peaked than either other component, with the molecular mass the least centrally concentrated. We indicate the radius beyond which CO(1--0) is detected at $<3\sigma$ with a dashed line in the middle panel; beyond this radius the profile is essentially entirely determined by the model fit parameters. The value of the H$_2$ surface density at this radius is very similar to typical surface densities at which the total gas mass transitions to largely neutral atomic gas (i.e. H{\scriptsize I}).
}\label{fig:physprofs}
\end{figure*}

The radial profiles of the stars, molecular gas, and obscured SFR are shown in Figure~\ref{fig:physprofs} under the assumptions of Section~\ref{conv}. While all three components peak in the galaxy center in absolute terms, the shape of the distributions clearly differs between them. From this Figure it is clear that the stellar mass is more centrally concentrated than either of the other components, with the molecular gas the least centrally peaked. This was to be expected from the S\'{e}rsic profile fits to each component, as the stellar light also showed the highest S\'{e}rsic index. Incidentally, the  value of the molecular gas surface density beyond which we no longer detect CO emission is similar to typical surface densities at which neutral atomic gas begins to dominate the total gas mass \citep[e.g.,][]{schruba11,krumholz13}. It is therefore possible that our observations have resolved the entirety of the molecular gas in this galaxy and that deeper observations would not result in detectable CO at larger radii. Additional evidence for this possibility comes from the fact that the B array configuration data alone recover the integrated CO line flux measured in the lower-resolution C array data (which did not spatially resolve the source and therefore captured the total CO emission).

\begin{figure*}[htb]
\includegraphics[width=\textwidth]{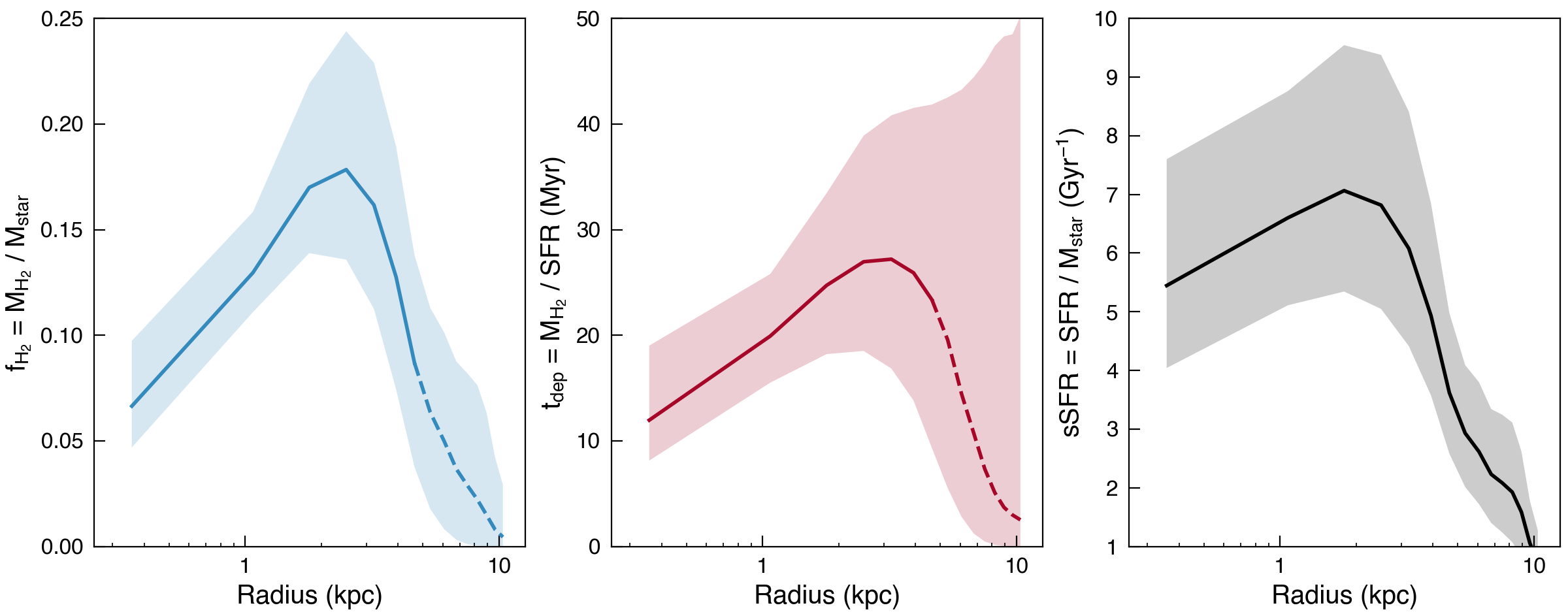}
\caption{
Radial profiles of \fht, \tdep, and sSFR in COSMOS\,27289. In all cases the 1$\sigma$ confidence intervals are shaded, determined from propagation of uncertainties in the radial profile fits. We indicate the radius beyond which CO(1--0) is detected at $<3\sigma$ with a dashed line in the left two panels; beyond this radius the profiles are mostly determined by the model fit parameters. We find lower values of both \fht and \tdep in the central regions of the galaxy, peaking at radii of $\sim2-3$\,kpc. The profile of sSFR tentatively follows the same behavior with large uncertainties, rapidly falling at radii $\gtrsim2-3$\,kpc.
}\label{fig:fgas_tdep_profs}
\end{figure*}

Figure~\ref{fig:fgas_tdep_profs} shows the radial profiles of the gas fraction, depletion time, and specific SFR determined by combining the individual profiles of Figure~\ref{fig:physprofs}. As in the two-dimensional map, we again find a central suppression of the gas fraction, rising to a maximum at radii $\sim$2-3\,kpc before falling. At radii $\gtrsim$5\,kpc the CO emission is no longer significantly detected; beyond this point the profiles are essentially extrapolations from the S\'{e}rsic models. We note that this radius is beyond the radius where \fht peaks; the peak and subsequent decline at larger radii is real. 

We see a similar trend in the radial profile of the depletion time \tdep, though with significantly larger uncertainties arising from the low resolution of the ALMA data. We find extremely short depletion times in the central regions of COSMOS\,27289, $\sim$10-15\,Myr, rising by a factor of 2 at radii of 3\,kpc. At even larger radii, the low resolution and low S/N of the data prevent robust constraints on the depletion time. Nevertheless, this Figure does indicate that quenching of the central kiloparsecs (and indeed, the galaxy as a whole) is imminent, as the galaxy will be unable to sustain its current rapid star formation without substantial gas accretion.

Finally, we see weak evidence that the specific SFR follows the same behavior as \fht and \tdep, perhaps slightly rising or remaining flat to radii of 2--3\,kpc before falling rapidly. In agreement with the radial gas fraction profile, this indicates that most of the mass buildup has transitioned out of the galaxy's central regions to larger radii. This is qualitatively similar to the behavior seen in larger samples of extended SFGs at similar redshifts, taken as evidence for inside-out growth and the formation of stellar bulges at the centers of galaxies \citep[e.g.][]{tacchella15,nelson16}.

Based on $\sim$0.14'' resolution ALMA observations of CO(8--7), \citet{barro17} argue for a high central gas fraction in a $z\sim2.3$ compact SFG, in apparent contrast to the central depletion we observe. However, CO(8--7) requires much higher gas temperatures and densities to significantly populate such high energy levels compared to the ground state CO(1--0) transition we observe. Gas with such high densities is generally also star-forming gas \citep[e.g.][]{greve14}. Thus, it may be more accurate to say that the dense, star-forming gas is concentrated in the galaxy studied by Barro \etal, in agreement with the centrally-concentrated SFR profiles we and others have observed. We do not consider our results to be in contradiction with those of Barro \etal, because our observations more faithfully trace the full molecular reservoir as opposed to only the very highly excited gas.

\subsection{Impact of Assumptions in Physical Conversion Factors} \label{assume}

It is worth considering how our conclusions could be influenced by the simplicity of our assumptions in converting observed to physical quantities. In general nothing is known about how the stellar mass-to-light ratio, CO-H$_2$ conversion factor, or dust flux density to SFR varies specifically in compact galaxies such as COSMOS\,27289. Here we outline general trends in these quantities seen in other galaxies in an attempt to illustrate the likely direction of these variations with galactocentric radius, if not their absolute magnitudes. 

First, the stellar mass-to-light ratio \ML is known to vary radially in galaxies. It would be quite difficult to quantify this effect in COSMOS\,27289 due to the high dust obscuration in this system and need for high-resolution imaging at longer infrared wavelengths. If the dust obscuration is moderate and not optically thick, then rest-frame optical color gradients are correlated with \ML; galaxies that are redder in their centers than the outskirts show falling \ML, and vice versa \citep[e.g.][]{zibetti09,wuyts12}. Equivalently, in a given band, the half-\textit{mass} radius will be smaller than the half-\textit{light} radius. Negative \ML gradients are far more commonly observed than the converse, although there is some evidence this effect becomes negligible at $z\gtrsim2$ \citep{suess19}. The typical magnitude of this variation is roughly a factor of 2 change in \ML over the full extent of galaxies. On the other hand if the obscuration is very high, then some amount of mass may be unaccounted for entirely by rest-UV/optical imaging. This would also imply a more centrally concentrated mass profile compared to our assumption, strengthening the central depletion of the gas fraction we observe.

Second, we also expect that our assumption of a single CO-H$_2$ conversion factor is too simplistic. We have assumed a value of $\alphaco = 1$. As we are mostly interested in the relative, rather than absolute, distribution of the molecular gas with respect to other properties, a single global revision of the value of \alphaco has no effect on our qualitative conclusions. However, we expect that the value of \alphaco is not uniform across the galaxy.  Observations of nearby galaxies show that the CO-H$_2$ conversion factor tends to be lower in the central regions before flattening at large radii by about a factor of two on average \citep{sandstrom13}. This is also in agreement with the observation that \alphaco is correlated with \lfir (or SFR; e.g. \citealt{spilker15}) and the fact that the SFR is centrally concentrated in COSMOS\,27289.  For a given CO luminosity, we expect \textit{lower} H$_2$ masses in the central regions than the outskirts due to a radially-varying \alphaco. In comparison to our assumption of a single CO-H$_2$ conversion factor, we thus expect that the true distribution of H$_2$ is \textit{less} centrally concentrated, again exaggerating the magnitude of the central gas depletion and rising depletion time with radius.

Finally, in regards to the conversion between rest-frame 270\,\um emission and SFR, we note again that the obscured star formation in COSMOS\,27289 completely dominates over the unobscured star formation ($>$98\% in the 3D-HST catalogs based on 24\,\um photometry and from our own re-analysis of archival \textit{Herschel} data). Using a tracer of the obscured star formation as a proxy for the total SFR is therefore of little concern. Using a single far-IR photometric point as a proxy for \lfir, and then the obscured SFR, is equivalent to assuming a single far-IR spectral template across the galaxy. If parameterized with a standard modified blackbody function, this would be equivalent to assuming a single dust temperature across the galaxy. Observations of nearby galaxies and simulations both indicate that negative temperature gradients are generally expected \citep[e.g.,][]{engelbracht10,casasola17,liang19}. The total dust luminosity has a steep dependence on temperature, and the dust in COSMOS\,27289 is most likely warmer in the central regions than the outskirts. For a given 270\,\um flux density, therefore, we expect \textit{higher} SFRs in the central regions than the outskirts due to the effect of dust temperature gradients. While we cannot quantify the magnitude of this effect with the data in hand, we expect that the true radial distribution of the SFR is \textit{more} centrally concentrated than our simple assumption indicates, exaggerating the trend we observe of rising depletion time with radius.

In summary, the probable radial variations in conversion factors to physical units in each case serve to accentuate the central low gas fraction and short depletion time we observe. The stellar \ML ratio likely decreases with radius, resulting in a more centrally concentrated stellar mass profile compared to our simple assumption. The CO-H$_2$ conversion factor likely rises to larger radii, resulting in less molecular gas in the center. The dust temperature probably decreases towards the galaxy outskirts, resulting in more IR luminosity and higher obscured SFR in the center compared to the outskirts. These three effects would further lower the gas fraction and shorten the gas depletion time in the galaxy center, resulting in a more extreme inside-out quenching scenario than our simple assumptions indicate.

\subsection{Inside-Out Quenching and\\Comparison to Simulations} \label{results:sim}

COSMOS\,27289 was selected as a galaxy thought to be undergoing rapid transformation from a highly star-forming galaxy to a $z\sim2$ quiescent object, with mass and structural properties very similar to the latter population \citep{barro13,barro14,vandokkum15}. Subsequent observations demonstrated that this galaxy is remarkably gas-poor compared to most star-forming galaxies at this epoch, consistent with a transition to quiescence on short timescales \citepalias{spilker16b}.

Our VLA observations offer a more detailed look at the quenching process in this galaxy. Taken at face value, both the map of the gas fraction and the radial profiles of \fht and \tdep are in excellent agreement with expectations for an inside-out quenching scenario, in which star formation ceases in the center of the galaxy before the outer regions. The low central gas fraction (with accompanying very high stellar mass surface density) implies that  COSMOS\,27289 formed through a strong central starburst that depleted the gas-rich central regions, converting a large fraction of the baryonic mass to stars. We have likely observed this object near the tail-end of this starburst -- the galaxy-integrated SFR is still quite high, but the remaining molecular fuel now plays its most important role at radii $\sim2-4$\,kpc. 

The radial distribution of the depletion time indicates that the central kiloparsecs of COSMOS\,27289 will exhaust the remaining molecular gas on extremely short timescales, $\sim10-15$\,Myr.\footnote{Or, more likely, the SFR will decline rapidly in concert with the diminishing gas reservoir, resulting in a longer time to total gas depletion.} In the absence of further gas accretion or migration from the outskirts into the central regions, therefore, only a minuscule further amount of in situ stellar buildup can occur. The radial gradient in \tdep will accentuate the central suppression of \fht on short timescales. The gas at larger radii can likely remain for a slightly longer period of time (though still very short compared to typical galaxy-integrated depletion times; \citealt{tacconi18}).  We note that a similar qualitative trend of rising \tdep with galactocentric radius is also seen in local spiral and dwarf galaxies but is not considered evidence of inside-out quenching because the depletion times are more than two orders of magnitude longer in the nearby objects \citep{leroy08}. It is much more likely that the gas in the central regions can be replenished on Gyr timescales in nearby galaxies compared to the 10\,Myr timescale we find for COSMOS\,27289.

The structure we observe in the gas fraction and depletion time bears some resemblance to similar patterns derived from hydrodynamical simulations of massive galaxies during quenching. Multiple simulation groups have seen ring-like structures in the gas and gas fraction distributions of galaxies at times when they are on the verge of a rapid decrease in SFR \citep[e.g.,][]{zolotov15,tacchella16,xma17}. In the simulations, this configuration develops as a result of powerful feedback and outflows that disrupt and evacuate the central regions of the galaxy of gas. These effects are weaker further from the central starburst, allowing the gas to remain relatively unscathed at larger radii. The fact that these ring-like structures are observed in simulations with different prescriptions for, e.g., supernova and AGN feedback and star formation, may imply that these features are a generic result of powerful central starbursts.

We perform a quantitative comparison to one of these simulations. We use the simulated outputs from \citet{tacchella16}, specifically of galaxies classified in their `quenching' phase according to those authors. We calculate the average radial gas fraction profile using the 12 simulated galaxies with stellar masses $\log\Mstar/\Msol>10.2$ at $z=2$, which quench at $z\sim1$--2 and have a mean stellar effective radius $\approx1.8$\,kpc. Because this simulation does not distinguish between gas that is specifically cold and molecular and gas in other forms, we estimate the molecular fraction using a crude analytic approximation from \citet{krumholz13}; this estimator has been previously used in similar fashion in other cosmological simulations \citep{lagos15}. As COSMOS\,27289 is both massive and compact, we apply the \citet{krumholz13} results assuming solar metallicity and high galactic stellar density and neglect any contribution from hot (non-molecular) gas. In other words, we assume that all of the gas is molecular at high column densities, transitioning to atomic at low column densities $\lesssim$30\,\Msol\,pc$^{-2}$. We note that this `correction' for the molecular fraction has no influence in the center of the galaxy where the surface densities are high; instead it merely serves to truncate the molecular gas at large radii where the surface densities are much lower than the typical threshold for molecule formation \citep[e.g.,][]{schruba11}.

\begin{figure}[htb]
\includegraphics[width=\columnwidth]{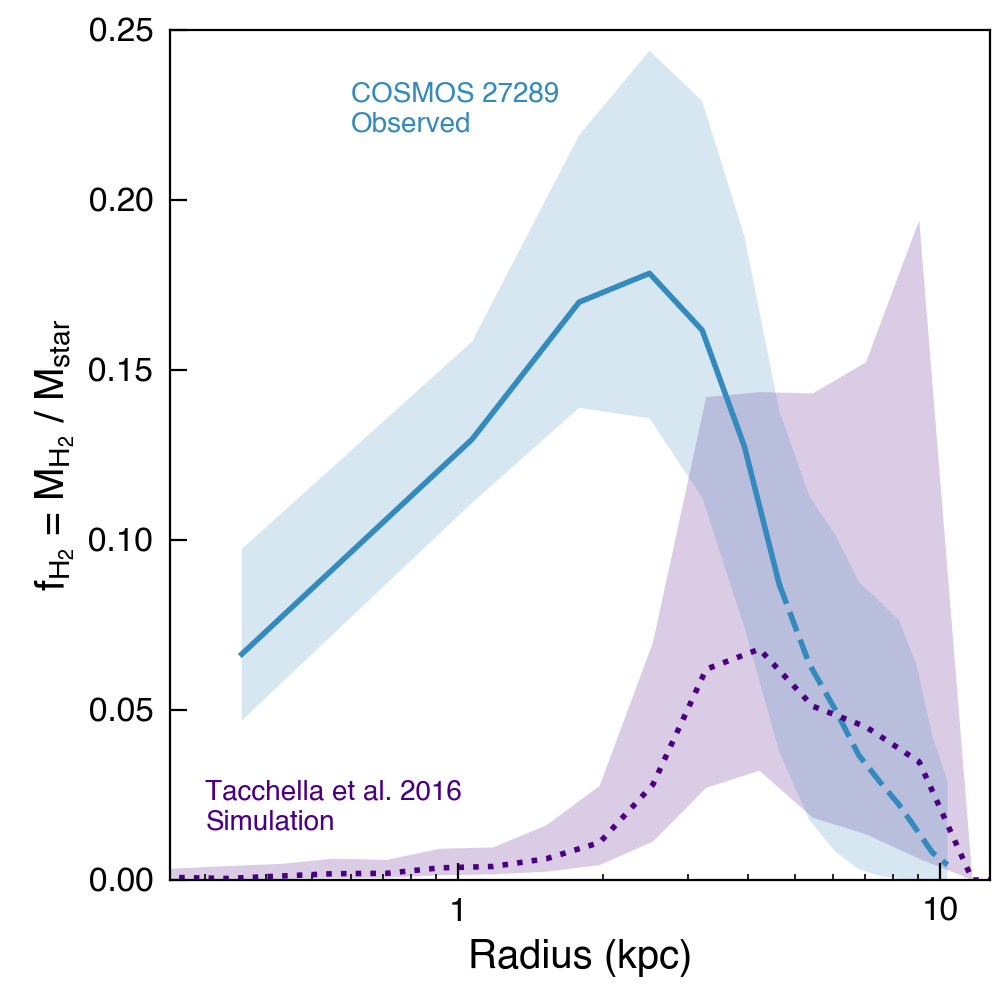}
\caption{
Comparison of the radial profile of the molecular gas fraction of COSMOS\,27289 with simulated massive galaxies undergoing quenching from \citet{tacchella16}. We find reasonable qualitative and quantitative agreement between this simulation and our observations, given the large uncertainties present in both.
}\label{fig:fgas_prof_t16}
\end{figure}

In Figure~\ref{fig:fgas_prof_t16} we show the radial profile of the gas fraction derived from the \citet{tacchella16} simulations and our observed profile of COSMOS\,27289. While the simulation shows an overall too-low normalization of \fht and a stronger-than-observed central suppression of the molecular gas, the qualitative agreement is relatively good. Compared to our observations, the simulated \fht profile peaks at larger radii $\sim$3-8\,kpc, about a factor of two larger than the observed peak in COSMOS\,27289. Inspection of Fig.~7 of \citet{tacchella16} indicates that this is mostly due to the fact that the simulated gas profile is significantly shallower than our observation indicates. 

Given the very short relevant timescales involved, it is possible that COSMOS\,27289 may come into better agreement with the simulated galaxies in $\sim$20\,Myr. Conversely, the simulated galaxies may have been in better agreement with our observations some short time before the simulation snapshots used to construct Figure~\ref{fig:fgas_prof_t16}, which are only saved every $\sim$150\,Myr at $z\sim2.3$. Given this limitation, we find that the quantitative comparison between our observations and this simulation suite is encouraging. 

\subsection{Comparison to Probable Progenitor and Descendant Populations} \label{results:context}

Our results demonstrate that COSMOS\,27289 is indeed likely to cease star formation on short timescales, and that this quenching will occur in the center of the galaxy first before proceeding to the outer regions. At that point the galaxy will appear remarkably similar to $z\sim2$ quiescent galaxies in structure. Here we briefly comment on how galaxies like COSMOS\,27289 may connect with probable progenitor and descendant populations of galaxies.

First, several authors have argued in favor of a connection between high-redshift IR-luminous dusty, star-forming galaxies (DSFGs) and early quiescent objects. These DSFGs are typified by very rapid star formation, high gas fractions, and short depletion times \citep[e.g.,][]{carilli13,casey14,aravena16}, with star formation and molecular gas reservoirs extended on scales of a few to 10\,kpc \citep[e.g.,][]{ivison11,ivison13,spilker15}. Evidence for the connection between DSFGs and early quiescent galaxies comes from a variety of arguments, including mass and number density arguments \citep[e.g.,][]{ivison98,straatman14}, clustering arguments \citep[e.g.,][]{williams11,hickox12}, and structural arguments \citep[e.g.,][]{toft14,spilker16}. In comparison to these objects, the mass, SFR, and depletion time of COSMOS\,27289 are comparable, but \fht is a factor of $\sim3-5$ lower than the gas-rich DSFGs. The short depletion times and low gas masses of compact SFGs imply that they are likely near the end of their rapid growth phase, perhaps shortly following an IR-luminous period. Finally, the central suppression of molecular gas and radial gradient in \tdep are produced in simulations due to the consequences of powerful galactic feedback processes that rapidly halt rapid star formation. Evidence for such feedback has recently been observed in a $z\sim5$ DSFG, which implies that the large molecular gas reservoirs of these galaxies can be rapidly depleted due to high-velocity molecular outflows \citep{spilker18b}. From a variety of perspectives, then, high-redshift DSFGs indeed appear to be plausible immediate progenitors of compact SFGs like COSMOS\,27289.

The short values of \tdep in compact SFGs imply that their current high SFRs must decline rapidly, ending the current burst of star formation. In this event, compact SFGs should soon present stellar spectra similar to other post-starburst galaxies selected at high redshifts. At low redshifts, post-starburst galaxies tend to have longer-than-expected depletion times \citep{french15}, though the `bursts' in these objects contribute only a minor amount to the existing stellar mass. In these galaxies, the depletion time is at a minimum at the conclusion of the burst, becoming two orders of magnitude longer over the ensuing $\approx$300\,Myr as the dense, star-forming gas is disrupted \citep{li19}. This scenario is in qualitative agreement with our results for COSMOS\,27289. At $z\sim2$, \citet{wild16} find that photometrically-selected post-starburst galaxies can account for the entirety of the massive quiescent population assuming the post-starburst features are observable for $\sim$250\,Myr.  These authors suggest that $z>2$ post-starbursts originated through both rapid assembly and rapid quenching, in agreement with our observations of COSMOS\,27289 and the discussion in the previous paragraph, while lower-$z$ objects originated through rapid quenching of more long-lived SFGs (see also \citealt{wu18b,belli19}). The molecular gas contents of post-starbursts outside the local universe are essentially unconstrained. \citet{suess17} present observations of two massive post-starburst galaxies at $z\sim0.7$, finding molecular gas masses that yield low gas fractions but depletion times of several Gyr, two orders of magnitude longer than our measurement of COSMOS\,27289.  Given its still rapid SFR, it may be that COSMOS\,27289 will be more completely depleted of molecular gas than either of those lower-redshift post-starbursts; ALMA observations of $z\sim2$ post-starbursts are required to make a more direct comparison.

Finally, if compact SFGs like COSMOS\,27289 will become quiescent on short timescales, it is worth briefly considering whether the available gas measurements of both populations are consistent with this picture. In the local universe, it has long been recognized that massive elliptical galaxies tend to be extremely gas-poor \citep[e.g.,][]{young11,davis19}. Far fewer constraints are available at higher redshifts due to the faintness of the gas tracers, which require the sensitivity of ALMA to detect. \citet{spilker18a} presented CO(2--1) observations of 8 massive and passive galaxies at $z\sim0.7$, in which we found very low gas fractions but also relatively short gas depletion times $\lesssim$1\,Gyr. This picture largely continues to $z\sim1.5$ \citep{sargent15,rudnick17,hayashi18,bezanson19}, where a total sample of four quiescent galaxies all exhibit strikingly low gas fractions (\tdep is more difficult to constrain as reliable SFRs become more challenging to measure). In summary, these observations of older quiescent galaxies all point to very efficient molecular gas depletion, in agreement with our current understanding of the future evolution of COSMOS\,27289 following its imminent depletion of molecular gas.

\section{Conclusions} \label{conclusions}

We have observed CO(1--0) emission at high spatial resolution in COSMOS\,27289, a compact SFG at $z=2.234$, and combine these new observations with archival data from HST and ALMA. Our previous lower-resolution VLA observations of this source revealed a very low molecular gas fraction and short gas depletion time consistent with a rapid transition to quiescence. From its morphology and dynamical arguments we determine that this galaxy must be oriented nearly face-on, allowing a relatively straightforward view of the radial distributions of the molecular gas, stellar mass, and obscured star formation. Multiple analysis techniques reveal that the gas fraction approximately doubles from $\sim0.07$ in the central 1--2\,kpc of this galaxy to its maximum at radii $\sim$2-4\,kpc. We find weaker evidence for a similar trend in the gas depletion time, doubling from 10--15\,Myr in the center to larger radii. Importantly, while we have made very simple assumptions to translate between observed and physical quantities, the probable radial variations in the stellar mass-to-light ratio, the CO--H$_2$ conversion factor, and light-weighted dust temperature would all serve to accentuate the central suppression of \fht and \tdep. 

Our results are in agreement with expectations for an inside-out quenching scenario, in which the centers of galaxies cease star formation before the outskirts, the first time this has been shown using molecular gas observations. Our observations provide the most direct evidence to date for inside-out quenching, supplementing previous studies focused on the radial distribution of star formation and stellar mass \citep[e.g.,][]{tacchella15}. High-resolution imaging of the molecular gas improves on such studies, as the results are immune to the effects of increased dust obscuration in the centers of galaxies. A comparison to hydrodynamical simulations of massive galaxy formation shows good qualitative and reasonable quantitative agreement \citep{tacchella16}, supporting a scenario by which the central regions of galaxies form in a rapid starburst that rapidly quenches due to gas consumption and galactic feedback. 

This case study targeted the only compact SFG detected in CO(1--0) emission thus far. Targeting ground-state CO emission avoids uncertainties present in other methods of molecular gas observations such as the CO excitation and mass-weighted dust temperature, at the cost of a not-insignificant observational investment. Future progress will likely require a variety of approaches, including observations of the dust continuum at multiple frequencies to determine the effects of dust temperature variations, observations of higher-J CO transitions that are significantly brighter than the ground-state transition, and/or observations of neutral carbon emission as a tracer of the molecular ISM \citep{popping17}. A larger sample of objects will also facilitate a more comprehensive picture of the role that compact SFGs such as COSMOS\,27289 play in the timeline of the evolution of massive galaxies, from higher-redshift progenitors to lower-redshift descendants.

\acknowledgements{
We thank the referee for a constructive report that improved the quality and clarity of the paper.
We thank Sandro Tacchella for providing simulation data products and for helpful discussions.
JSS thanks the McDonald Observatory at the University of Texas at Austin for support through a Harlan J. Smith Fellowship.
CCW acknowledges support from the National Science Foundation Astronomy and Astrophysics Fellowship grant AST-1701546.
This paper makes use of the following ALMA data: ADS/JAO.ALMA\#2015.1.00137.S. ALMA is a partnership of ESO (representing its member states), NSF (USA) and NINS (Japan), together with NRC (Canada), MOST and ASIAA (Taiwan), and KASI (Republic of Korea), in cooperation with the Republic of Chile. The Joint ALMA Observatory is operated by ESO, AUI/NRAO and NAOJ. The National Radio Astronomy Observatory is a facility of the National Science Foundation operated under cooperative agreement by Associated Universities, Inc.
This work is based on observations taken by the 3D-HST Treasury Program (GO 12177 and 12328) with the NASA/ESA HST, which is operated by the Association of Universities for Research in Astronomy, Inc., under NASA contract NAS5-26555.
This research has made use of NASA's Astrophysics Data System.
}

\facilities{VLA, HST (WFC3), ALMA}

\bibliographystyle{apj}

\end{document}